# Nonlinear multi-frequency phonon lasers with active levitated optomechanics

Tengfang Kuang[1,5], Ran Huang[2,5], Wei Xiong[1], Yunlan Zuo[2], Xiang Han[1], Franco Nori[3], Cheng-Wei Qiu[4,*], Hui Luo[1,*], Hui Jing[2,*], Guangzong, Xiao[1,*]

[1] College of Advanced Interdisciplinary Studies, NUDT, Changsha Hunan, 410073, China

[2] Department of Physics and Synergetic Innovation Center for Quantum Effects and Applications, Hunan Normal University, Changsha 410081, China

[3] Theoretical Quantum Physics Laboratory, RIKEN Cluster for Pioneering Research, Wako-shi, Saitama 351-0198, Japan

[4] Department of Electrical and Computer Engineering, National University of Singapore, 4 Engineering Drive 3, Singapore, 117576, Singapore

[5] These authors contribute equally to this work

*email: chengwei.qiu@nus.edu.sg; luohui.luo@163.com; jinghui73@foxmail.com; xiaoguangzong@nudt.edu.cn.

**Phonon lasers, exploiting coherent amplifications of phonons, have been a cornerstone for exploring nonlinear phononics, imaging nanomaterial structures, and operating phononic devices. Very recently, by levitating a nanosphere in an optical tweezer, a single-mode phonon laser governed by dispersive optomechanical coupling has been demonstrated, assisted by alternating mechanical nonlinear cooling and linear heating. Such levitated optomechanical (LOM) devices, with minimal noises in high vacuum, can allow flexible control of large-mass objects without any internal discrete energy levels. However, untill now, it is still elusive to realize phonon lasing with levitated microscale objects, due to much stronger optical scattering losses. Here, by employing a $Yb^{3+}$-doped active system, we report the first experiment on nonlinear multi-frequency phonon lasers with a micro-size sphere governed instead by dissipative LOM coupling. In this work, active gain plays a key role since not only 3-order enhancement can be achieved for the amplitude of the fundamental-mode phonon lasing, compared with the passive device, but also nonlinear mechanical harmonics can emerge spontaneously above the lasing threshold. Furthermore, for the first time, coherent correlations of phonons are observed for both the fundamental mode and its**

**harmonics. Our work drives the field of LOM technology into a new regime where it becomes promising to engineer collective motional properties of typical micro-size objects, such as atmospheric particulates and living cells, for a wide range of applications in e.g., acoustic sensing, gravimetry, and inertial navigation.**

Conventional optomechanical systems rely on fixed frames to support mechanical elements, leading to unavoidable energy dissipation and thermal loading[1]. LOM system, i.e., controlling motions of levitated objects with optical forces[2], have provided unique advantages[3], such as fundamental minima of damping and noises, the possibility for levitating complex objects, as well as high degree of control over both conservative dynamics and coupling to the environment. These advantages are of significance for both fundamental studies of non-equilibrium physics and applications in metrology[4-11]. In recent years, remarkable achievements have been witnessed in LOM systems[12-14], such as the realizations of motional ground-state cooling[15,16], room-temperature strong coupling [17], or ultrahigh-precision torque sensing[18], to name only a few. In a very recent work[19], a phonon laser or coherent amplification of phonons, the quanta of vibrations, was demonstrated for a levitated nanosphere, based on dispersive LOM coupling, in which the optical resonance frequency is modulated by mechanical motion. This work offers exciting opportunities of exploring the boundary of classical and quantum worlds with levitated macroscopic object[16,20,21], as well as making various levitated sensors[22,23]. Nevertheless, sophisticated feedback controls based on electronic loops[19] are needed to provide both nonlinear cooling and linear heating, and only a single-mode phonon laser was observed, without any evidence of nonlinear mechanical harmonics.

Besides LOM systems, phonon lasers have also been built by using semiconductor superlattices[24], nanomagnets[25], ions[26], and nanomechanical[27] or electromechanical[28] devices. These coherent sound sources, with shorter wavelength of operation than that of a photon laser of the same frequency, are indispensable in steering phonon chips[30], improving the resolution of motional sensors[31], and exploring new effects of phonons[32-34]. However, as far as we know, the ability of achieving multi-frequency phonon lasers with micro-size levitated objects, has not been reported. This ability can be the first step for many important applications such as multi-frequency motional sensors, exceptional point optomechanics[32,33], and topological sound-wave control[35].

In this Letter, we develop a strategy to achieve nonlinear phonon lasers for a levitated object at microscales by utilizing an active LOM system. We show that in such a system, dissipative LOM coupling[36-39] can be significantly enhanced by introducing an optical gain, thus leading to not only efficient output of fundamental-mode phonon lasing, but also spontaneous emergence of mechanical harmonics. The active gain plays a key role in our work since for passive systems, only thermal phonons exist, and no phonon lasing can appear. To steer this system from a chaotic regime into a phonon lasing regime, an optical gain is used to increase the photon lifetime and thus enhance the LOM coupling; as a result, *three-order enhancement* in the power spectrum of the fundamental-mode phonons is achieved, with *40-fold narrowing* in its linewidth. More importantly, above the lasing threshold, we observe nonlinear harmonics with double and triple mechanical frequencies, as clear evidence of gain-enhanced nonlinearity in an active LOM system.

We stress that our work is the first experiment on phonon lasing with a micro-object which is 3 to 4 orders larger in size or mass than a nano-sphere[19]. This ability, never achieved in previous works[24-28], is an important step towards applications based on coherent motional control of a wide range of typical micro-size objects. Also, our work is the first demonstration of mechanical harmonics accompanying phonon lasing, with evidence of threshold features and high-order correlations. The key role of active gain in enhancing nonlinear LOM effects, as far as we know, is also the first example reported untill now, opening up the door to engineer and utilize more nonlinear phononic effects in LOM systems, and to achieve exciting goals well beyond the reach of passive systems, e.g., parity-time-symmetric or gain-enhanced LOM sensing. Finally, our work is the first example showing dissipative optomechanical coupling in LOM architectures, and compared to dispersive ones, dissipative LOM systems are expected to have unique advantages in acoustic metrology[40].

Our experimental platform includes an active optical cavity and a dual-beam optical tweezer for trapping a single microsphere (Fig. 1a). The active cavity works for all three translational degrees of freedom, well tuneable along both the longitudinal and vertical directions. Its quality factor is $10^6$ but can be enhanced to $10^9$ by applying a $Yb^{3+}$-doped gain fibre. The photon lifetime is thus increased to more than $1\ \mu s$, enhancing coherent vibrational amplification of the sphere. Technically, the optical gain is achieved simply by using a Yb-doped fibre with a pumping laser at 976 nm, and no other limitation exists. Compared to the cavity-free work[19], there is no need to design and carefully control both nonlinear cooling and linear heating of the mechanical motion with complicated electronic loops or algorithms[19] (see Supplementary Section 1 for details).

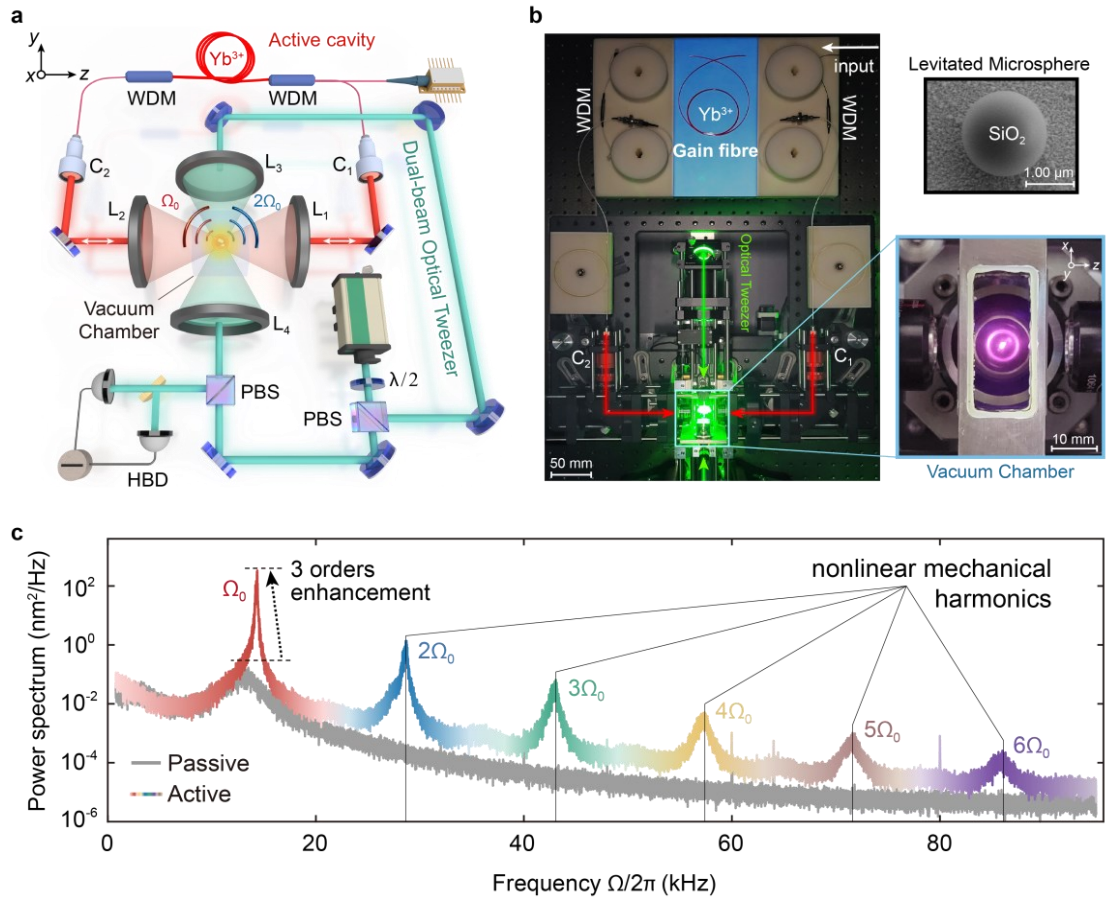

**Fig. 1 Experimental overview. a,** Schematic diagram of the active LOM system, including an active optical cavity (red) and a dual-beam optical tweezer (green). WDM, wavelength division multiplexer; $C_1$, $C_2$, collimators; $L_1 \sim L_4$, lenses; PBS, polarizing beam splitter; HBD, heterodyne balanced detection. **b,** Photograph of the active LOM setup with the vacuum chamber, and the scanning electron microscope image of the levitated microsphere. **c,** Measured power spectra of phonons in an active cavity (coloured curve) and a passive cavity (grey curve).

As shown in Fig. 1b, essentially different features can be observed in the power spectrums of phonons for the cases with or without the gain, i.e., only thermal phonons can exist in the passive case, while a 3-order-of-magnitude enhancement can be achieved in the active case for the fundamental mode with the mechanical frequency $\Omega_0 = 14.4\,\text{kHz}$. In particular, nonlinear mechanical harmonics with frequencies $2\Omega_0$, $3\Omega_0$, ... emerge spontaneously in the spectrum, enabling the observation of tuneable multiple-frequency phonon lasers (see also Supplementary Section 1). Besides the role of gain in achieving phonon lasers, we expect that active LOM systems can also serve as an important tool for studying, e.g., parity-time-symmetric optomechanics[32] or gain-enhanced sound sensing[41].

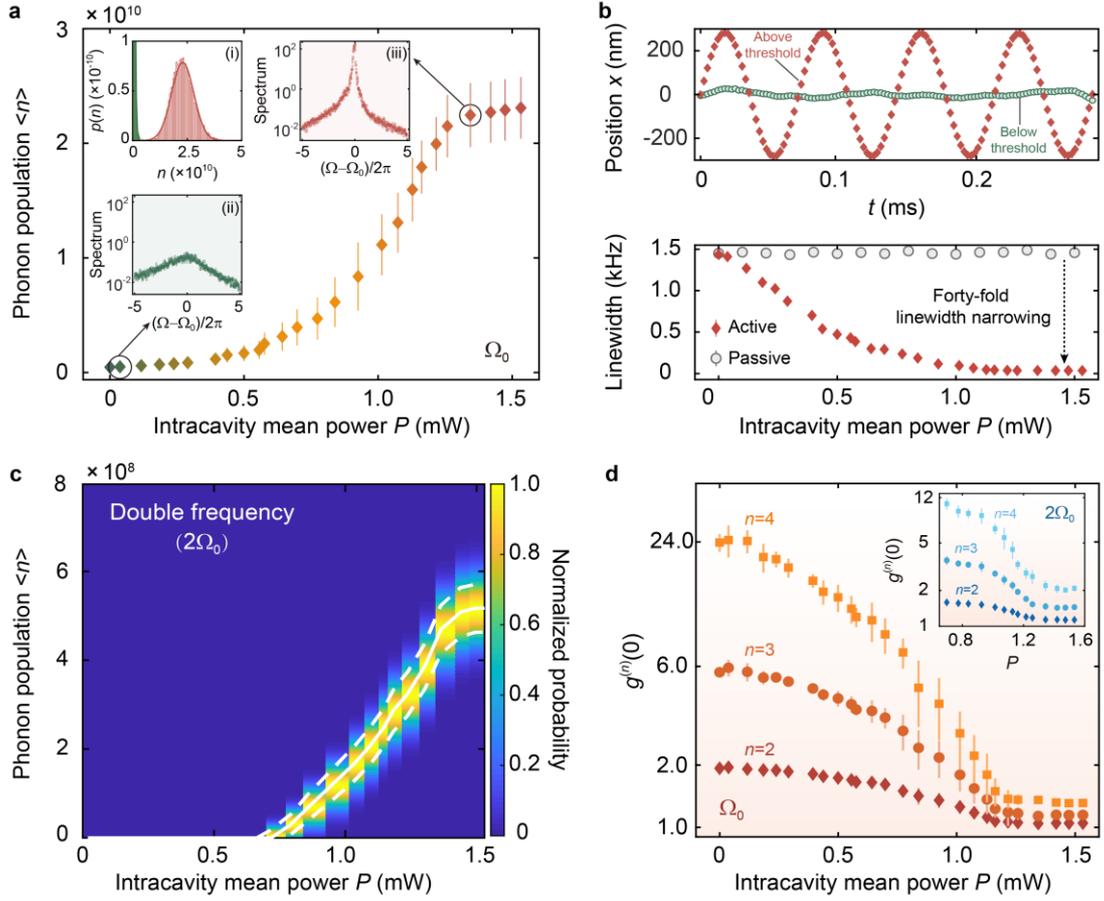

**Fig. 2 Experimental results of phonon lasing with nonlinear harmonics. a,** Phonon populations with fundamental frequency $\Omega_0$ as a function of intracavity mean power $P$. The measured threshold power is in good agreement with the theoretical prediction. The insets show (i) the phonon probability distributions, and (ii, iii) the phonon power spectra below or above the threshold. **b,** The measured oscillation dynamics (upper panel) and linewidths (lower panel) of the fundamental mode by tracing the $SiO_2$ micro-sphere. **c,** Threshold behaviour of the phonon laser with double frequency $2\Omega_0$ (white solid curve). The white dashed curves represent ±1 s.d. of each measurement, consisting of $5\times10^5$ samples. The normalized phonon probability distribution is displayed in colour. **d,** Measured $n$th-order phonon autocorrelations at zero-time delay $g^{(n)}(0)$ versus $P$ for $\Omega_0$. Inset: $g^{(n)}(0)$ for $2\Omega_0$. Error bars denote ±1 s.d. of each measurement, consisting of $5\times10^5$ samples.

To confirm the phonon lasing of the fundamental mode and its harmonics, we firstly measure the steady-state phonon population $\langle N \rangle = M\Omega_0 \langle x^2 \rangle / \hbar$, where $M$ is the mass of the levitated sphere, $\Omega_0$ is the oscillation frequency of the mode, $x$ is the centre of mass displacement of the sphere, and $\hbar$ is the reduced Planck's constant. Explicit signatures of lasing threshold are observed for the fundamental mode, as shown in Fig. 2a, by increasing the intracavity mean power $P$. The threshold value is $P_{\text{th}} = 0.39$ mW, which agrees well with theoretical calculations (Supplementary Section 2). The insets of Fig. 2a showcase a linewidth narrowing, accompanying the transition from thermal to coherent oscillations. Below the threshold, the oscillator features thermal dynamics with mean phonon number $4.63\times10^8$, and the phonon probability distribution is well described by the Boltzmann distribution. By surpassing the lasing threshold, the phonon number is greatly enhanced towards its saturation value $2.3 \times 10^{10}$ at $P = 1.3 \, \text{mW}$, also accompanied by a significant narrowing of linewidth. This is closely related to the fact that phonon lasing emerges in the system, resulting in the Gaussian distribution of the generated coherent phonons. By further increasing the power, the phonon population gradually approaches its saturation value.

Figure 2b further presents the experimental results of the dynamical behaviours of the micro-sphere. We find that significant motional amplifications of the micro-sphere emerge above the lasing threshold, with 40-fold improvement in linewidth narrowing, compared to the case without any gain. Clearly, in the passive case, the interaction between the light and the levitated microsphere is rather weak, due to the large optical loss, and the linewidth remains at about $1.5 \, \text{kHz}$. In contrast, in the active LOM system, the linewidth can approach as low as $0.03 \, \text{kHz}$ above the threshold.

Accompanying the giant enhancement of fundamental-mode phonon lasing, we also observe spontaneous emerging of nonlinear mechanical harmonics. We find that similar lasing features for the double-frequency mechanical mode can be achieved as shown in Fig. 2c, which as far as we know, have never been achieved in LOM systems[19]. It is also different from the multi-mode phonon laser shown in a flat membrane trapped in a Fabre-Perot cavity[42], in which mode competitions make it only possible to stimulate a single phonon mode into the lasing regime.

To further reveal the coherence of the phonon lasers, we study the *n*th-order phonon autocorrelation functions at zero-time delay $g^{(n)}(0)$ (see Supplementary Section 2.2). For the phonon laser in fundamental mode, we find that the *n*th-order correlations of the phonons satisfy $g_0^{(n)} = n!$ below the threshold $P_{\text{th}}$, demonstrating their thermal statistics. As $P$ exceeds $P_{\text{th}}$, $g^{(n)}(0)$ decreases and approaches 1, i.e., the phonon dynamics changes from the thermal state to the coherent state. Moreover, $g^{(n)}(0)$ approaches 1 for the harmonics with $2\Omega_0$ when operating in the lasing regime, as shown in the onset of Fig. 2d.

We remark that the origin of these nonlinear mechanical harmonics is the anharmonic optical potential produced by the optical-gain-enhanced nonlinearity. Without the gain, for a sphere with larger size than that used in Ref.[19], stronger scattering losses lead to a smaller cavity quality factor and thus weaker light-motion coupling. Therefore, we find that the intracavity power $P$ is independent of the $x$-position of the oscillator (see

Supplementary Section 3), while the optical force $F_{opt}$, relying on both of $P$ and $x$, responds linearly to the position $x$ (Fig. 3a). By measuring the phonon dynamics for spheres with different sizes (Fig. 3b), we also find that the linewidth of the fundamental mode is invariant for spheres smaller than 1.2 μm; the phonon lasing only emerges for spheres with larger sizes, and becomes stronger when the size is increased to 2 μm. For the active case, the intracavity power can be modulated by the mechanical position due to strong light-motion coupling, and thus we find a strongly nonlinear optical force, as shown in Fig. 3a. As a result, the double-frequency component emerges in the phonon power spectrum (Fig. 3c), with also a giant enhancement compared to the passive case with the same intracavity power $P = 1.3$ mW. Similar lasing features are also observed for the harmonics with the triple-frequency $3\Omega_0$ as shown in Fig. 3d, which is otherwise impossible in the absence of an optical gain.

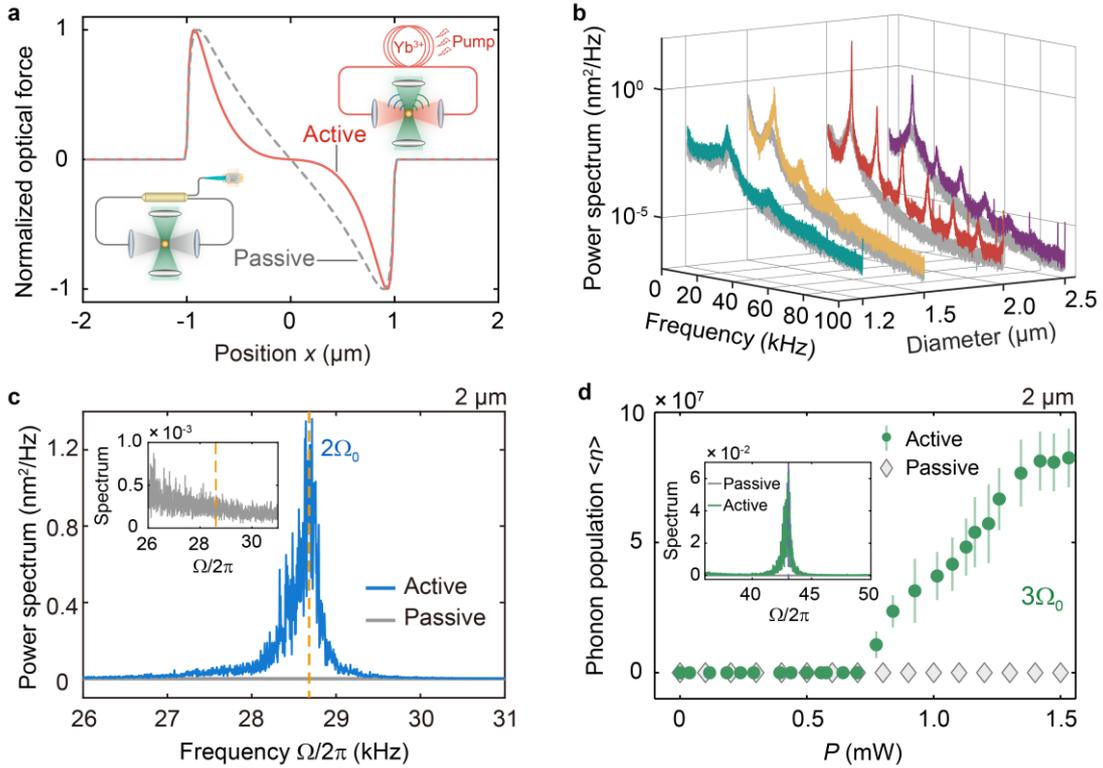

**Fig. 3 Gain-induced phonon lasing with double and triple frequencies. a,** Normalized optical force distributions for active (red solid curve) and passive (grey dashed curve) cases. The active case shows a strongly nonlinear distribution for the position $x$ between (−1, 1) μm, which is instead linear for the passive case. **b,** Power spectra for spheres of different sizes trapped in active (blue curve) or passive (grey curve) case, and **c,** Power spectra around $2\Omega_0$ for such two cases. Inset: 30 times magnified spectrum for better view. **d,** Phonon population with $3\Omega_0$ versus the intracavity mean power $P$. Phonon laser with $3\Omega_0$ (green dots) or thermal phonons (grey curve) can be clearly observed for different cases, with the corresponding power spectra shown in the Inset.

We note that, our results are essentially different from those in a very recent experiment on mechanical parametric amplification[43] which originates from the nonparabolicity of the optical potential, and typically dominates in the μm region for motional displacement (see also Ref.[44]). In contrast, in our work, the maximum displacement of the oscillator is ~300 nm, which is small enough and thus far away from the region of observing parametric amplification. In fact, in our work, the nonlinear optical force coming from dissipative LOM coupling exists even for smaller displacements (see Supplementary Section 1). Moreover, in the parametric amplification system, the nonparabolicity of the optical potential can lead to not only frequency multiplication, but also frequency shifts. When the oscillator experiences larger displacements, the absolute value of the optical force is a convex function, the slope of which (and thus the natural frequency of the oscillator) decreases, while in our work, the natural frequency increases above the lasing threshold. In addition, the origin of the nonlinearity of our work is fundamentally different from that in Ref.[19] (see Supplementary Section 1 for details). In our future work, by further combining with other existing techniques used in previous works[19,43], it is possible to probe different nonlinear mechanisms, allowing studies of more nonlinear LOM effects and additional flexible control of LOM devices.

In summary, we have experimentally reported nonlinear phonon lasers in an active LOM system. By introducing optical gain, we have realized a phonon laser on the fundamental mode with three-order-of-magnitude enhancement in the power spectrum, and forty-fold improvement in linewidth narrowing, without the need of any complicated external feedback control techniques. We also present unequivocal evidence of lasing threshold behaviour, and the phase transition from thermal to coherent phonons by measuring the phonon autocorrelations. More interestingly, for the first time, we observe nonlinear phonon lasers with multiple frequencies, resulting from the optical-gain-enhanced nonlinearity. As far as we know, this is the first observation of such nonlinear mechanical harmonics in LOM systems, which does not rely on the specific material or the shape of the oscillator[45,46]. We measured also correlations $g^{(n)}(0)$ of harmonic phonon lasing for the first time. These results push forward phonon lasers into the nonlinear regime and make many exciting applications more accessible, such as optomechanical combs[47], high-precision metrology, and non-classical state engineering. Our work opens up new perspectives for achieving levitated phonon devices with active LOM systems, and enables a wide range of applications such as quantum phononics, multi-frequency mechanical sensors, and high-precision acoustic frequency combs.

We stress that the purpose of our work is not to outperform dispersive phonon lasers; instead, it enriches the present toolbox of phonon lasing by confirming that, even for a micro-object containing as much as $\sim 10^{11}$ atoms, i.e., 4 orders of magnitude higher than that used in Ref.[19], is still possible to achieve collective mechanical amplifications and observe the accompaying nonlinear effects. In fact, our work can be combined with existing techniques from previous works, for future studies of more exciting projects, e.g., switch phonon lasing from dissipative-coupling-governed regime to dispersive-coupling-governed regime, transient LOM effects with comparable couplings, and the role of gain in operating nonlinear phonon lasers for sound-sensing applications.

Note added: When submitting our revised manuscript, a new experiment appeared[40], and in that work, different features of dispersive and dissipative couplings in sound sensing are compared, by using a different system, i.e., a suspended fibre coupled with an optical resonator.

## Methods

**Cavity alignment.** We mount the pumping laser to one collimator (Thorlabs, ZC618FC-B), and a power meter to another. The alignment is evaluated by the coupling coefficient from one collimator to another. Through adjusting the lenses and mirrors, the loss of the free-space optical path can be regulated to its lowest value (usually lower than 0.31 dB).

**Micro-sphere trapping.** The spheres was loaded to the trapping region by using an ultrasonic nebulizer, composed of an ultrasonic sheet metal with a great number of 5 μm holes distributed. It was trapped at atmospheric pressure. In most cases, a micro-sphere can be trapped within 30 s. Then, we reduced the pressure to the desired experimental level.

**Phonon population distribution.** The phonon population is obtained according to the relationship $\langle N \rangle = M\Omega_0 \langle x^2 \rangle / \hbar$, where $M$ is the mass of the levitated sphere, $\Omega_0$ is the oscillation frequency of the mechanical mode, $\langle x^2 \rangle$ is the mean square displacement of the sphere's centre-of-mass, and $\hbar$ is the reduced Planck's constant. Given a series of trajectories $x_i$ with $i = 1,\ldots L$, the distribution $\rho(n)$ is constructed as follows: First, start from $i = 1$, extract a finite dataset from $x_i$ with length $k$. Loop all datasets to calculate the phonon population as $N_i = \frac{M\Omega_0}{\hbar} \frac{1}{k} \sum_{j=i}^{k} (x_j - \bar{x})^2$. Second, determine the common maximum and minimum value of the phonon population variable, denoted by $N_{max}$ and $N_{min}$, and discretize the interval $[N_{max}, N_{min}]$ into $P$ bins as $N_\alpha = N_{min} + \frac{\alpha}{P}(N_{max} - N_{min})$, $\alpha = 0, 1, \ldots, P-1$. Then, use the $i$-th time series to calculate the frequency of data points falling in each subinterval, obtaining the distribution for the experiments $\rho(n)$.

**Data availability**

The data that support this article are available from the corresponding author upon reasonable request.

## Acknowledgements

This work is supported by the National Natural Science Foundation of China (Grants Nos. 61975237, 11904405, 11935006 and 11774086), the Science and Technology Innovation Program of Hunan Province (Grant No. 2020RC4047), Independent Scientific Research Project of NUDT (Grant No. ZZKY-YX-07-02), and Scientific Research Project of NUDT (Grant No. ZK20-14). We gratefully acknowledge the valuable assistance from Bin Luo at the BUPT, Şahin K. Özdemir at the PSU, Yafeng Jiao, Xunwei Xu at HNU, and Zijie Liu, Weiqin Zeng, and Xinlin Chen at NUDT.

## Author contributions

G.X. and H.J. conceived the idea. T.K. and G.X. designed the experiments. T.K., W.X. and X.H. performed the experiments and analysed the experimental data with the help of G.X. R.H. and T.K. performed the theoretical analysis and numerical simulations, guided by H.J. R.H., T.K. and Y.Z. wrote the manuscript with contributions from G.X., H.J., F.N., and C.W.Q. G.X., H.J. and H.L. supported the project.

## Competing interests

The authors declare no competing interests.

## Additional information

**Supplementary information** is available for this paper at http/